Title:

# The energy distribution structure and dynamic characteristics of energy release in electrostatic discharge process

Authors:

Qingming Liu[1], Huige Shao[1], Yunming Zhang[2]

Affiliation:

[1]State Key Laboratory of Explosion Science and Technology

Beijing Institute of Technology, Beijing, 100081, China

[2]Department of Fire Protection Engineering

Chinese People's Armed Police Force Academy

Corresponding author: Qingming Liu

Address: No. 5 South Zhongguancun Street, Haidian District, Beijing 100081, China

Telephone number: 86-10-68914261

Fax number: 86-10-68914287

E-mail address: qmliu@bit.edu.cn;

# The energy distribution structure and dynamic characteristics of energy release in electrostatic discharge process

Qingming Liu, HuigeShao, Yunming Zhang

Abstract:

The detail structure of energy output and the dynamic characteristics of electric spark discharge process have been studied to calculate the energy of electric spark induced plasma under different discharge condition accurately. By using spark discharge test system, a series of electric spark discharge experiments were conducted with the capacitor stored energy in the range of 10J, 100J, and 1000J, respectively. And the resistance of wire, switch, and plasma between electrodes were evaluated by different methods. An optimized method for electric resistance evaluation of the full discharge circuit, three poles switch and electric spark induced plasma during the discharge process was put forward. The electric energy consumed by wire, electric switch and electric spark induced plasma between electrodes were obtained by Joule's law. The structure of energy distribution and the dynamic process of energy release during the capacitor discharge process have been studied. Experiments results showed that, with the increase of capacitor released energy, the duration of discharge process becomes longer, and the energy of plasma accounts for more in the capacitor released energy. The dynamic resistance of plasma and three poles switch obtained by energy conversation law is more precise than that obtained by the parameters of electric current oscillation during the discharge process.

**Keywords**: electric spark, plasma, discharge characteristics, energy structure

**0 Introduction**

Electric spark is a phenomenon with thermal, luminous and acoustic energy when a high voltage is applied to gases, leading the ionization breakdown of gaseous medium[1]. Electric spark discharge is the main ignition mode in laboratory experiments and industrial application, such as gas or dust explosion and inter combustion engine, because the discharge process is ease to control and the electric energy is relatively concentrated[2]. And it also has a wide range of applications in aviation, military, metallurgy, chemical industry and everyday life. Spark energy is an important index to evaluate the ignition properties and hazardous characteristics of electric spark. Precise calculation and control of electric spark energy were required for effective ignition and safety of industry process. The method for electric spark energy calculation, the energy distribution and the dynamic characteristics of capacitor discharge process were of great important to the reliability and safety of industrial process.

The electric spark discharge process is very complicated as it is related to electrodes spacing, capacitance, resistance and many other factors. Maly et al.[3-6] divided the discharge process into four stages, pre-breakdown phase, breakdown phase, arc phase and glow phase. Usually the capacitor stored energy was taken as the electric spark energy. In practice, not all the energy stored in capacitor turns into spark energy, most of it is consumed by circuit. In order to calculate spark energy more accurately, a controllable ignition device was established. A serious of experiments on electric energy release during capacitor discharge process were conducted. The dynamic characteristics and the structure of electric energy distribution have been studied.

## 1 Experimental device and principle

### 1.1 Experimental device

The experimental system includes an electric spark generation system and a measurement system. The electric spark generation system consist of a high-voltage power, an energy storage capacitor bank, an three poles switch which is controlled by a trigger device, and a pair of tungsten electrodes. The measurement system consist of a TektronixDPO7254C digital oscilloscope, TektronixP6015A voltage probes and Pearson110A current monitors, as shown in Fig.1. The selection of components and technical parameters of the electric spark generation system can reference paper[7]. First the energy storage capacitor was charged by high voltage power, and then controlled by the three poles switch, the energy stored in the capacitor released through electric circuit and electric spark induced plasma between the tungsten electrodes. The energy stored in capacitor can be controlled by selecting the capacitor with different capacitance in the capacitor bank and the charging voltage. Electric energy release experiments were conducted with the stored energy in the range of 10J, 100J and 1000J respectively. The dynamic current of capacitor discharging process was monitored by a current sensor. And the dynamic voltage of capacitor and spark gap during the discharge process were measured by high voltage probes. The dynamic current and voltage can be used to evaluate the structure of energy distribution and the dynamic characteristics of capacitor discharge process. .

### 1.2 Evaluation methods for electric spark energy

In present study, the dynamic current and voltage were monitored by dynamic sensors of current and voltage respectively. And the data was recorded by a oscilloscope. A typical voltage history of capacitor and current history during the discharge process were shown in Fig.2. It can be found that the current and voltage are all subject to periodic damping oscillation. Up to now, there is not a uniform standard for electric spark energy evaluation. In Chinese standard GB/T16428—1996[8], International Electro technical Commission standard IEC 61241:1994 [9], European standard EN 13821:2002[10] and ASTM Standard E2019-03[11], the electric spark energy is evaluated by mathematical integration of *iu*, that is, $\mathrm{E_{spark}} = \int iu\mathrm{d}t$, where *u* and *i* are voltage and current histories respectively. With the residual energy of capacitor discharge process and the energy consumed by electric circuit considered, three different methods for spark energy calculation were analyzed and compared by Liu Qingming et. al[12]. It was found that every calculation method has its own advantages and disadvantages and suit for different conditions. Zhang Yunming et. al[7] found that the phase shift between current and voltage histories of spark gap resulting from sensors can't be neglected when the capacitor stored energy is large enough. So, the electric spark energy calculated by integration, $\int_0^t ui\, dt$, could lead to a considerable error. So they raised two other methods, Joule's integration method, that is $E_{spark} = \int_0^t i^2 R_{spark}\, dt$, and overall energy conservation method, that is $E_{spark} = E_{released} - \int_0^t i^2 R_{circuit}\, dt$ to calculate electric spark energy. Where, $E_{spark}$, and $E_{released}$ are electric spark energy and the capacitor released energy respectively during the discharge process. In ref[7] the resistance of electric induced plasma, $R_{spark}$, and the resistance of circuit, $R_{circuit}$, were evaluated by the dynamic parameters of current oscillation and attenuation. In this paper, an energy conservation based method for spark resistance evaluation was put forward.

Taking the residual energy into account, the energy released during the capacitor discharge process can be obtained by

$$E_{release} = \frac{1}{2}C(U_0^2 - U_1^2) \quad (1)$$

Where $C$ is the capacitance of energy storage capacitor, $U_0$ and $U_1$ are electric voltage of capacitor before and after discharge process, as shown in Fig. 2. From Fig. 2 we can see that the voltage is gradually decreased from a certain value, $U_0$, and finally remains constant value $U_1$.

Suppose the energy released is completely transformed into heat energy, the total resistance of the spark discharge process can be obtained by

$$R_{total} = \frac{C(U^2 - U_0^2)/2}{\int i^2 dt} \quad (2)$$

The time interval of integral is the duration of discharge process.

The resistance of short circuit, $R_{short}$, consist of wire resistance, $R_{wire}$, and switch resistance, $R_{switch}$, that is, $R_{short} = R_{wire} + R_{switch}$. $R_{wire}$ can be obtained by four wire electric resistance measure method, in our experiment circuit, $R_{wire}$ = 0.029 Ω. And short circuit resistance can be obtained by using the histories of voltage and current during energy release process of short circuit.

$$R_{short} = \frac{C(U_0'^2 - U_1'^2)}{\int i'^2 dt} \quad (3)$$

Where, $R_{short}$ is short-circuit resistance, $U_0'$ and $U_1'$ are voltage of energy storage capacitor before and after energy release process of short circuit, $i'$ is the dynamic current during energy release process of short circuit.

The resistance of electric spark induced plasma can be obtained by subscribe the short circuit resistance from the total resistance of the discharge circuit. By Joul's law, the electric spark energy can be expressed as

$$E_{spark} = \int i^2 dt \cdot (R_{total} - R_{short}) \tag{4}$$

So, the energy consumed by wire resistance is

$$E_{wire} = \int i^2 dt \cdot R_{wire} \tag{5}$$

And the energy consumed by switch can be expressed as

$$E_{switch} = \int i^2 dt \cdot (R_{short} - R_{wire}) \tag{6}$$

Using equations (4-6), the energy consumption by electric spark, wire and switch can be calculated, so the energy distribution structure of the released energy during capacitor discharge process can be obtained.

In above method, the capacitor released energy is divide into electric spark consumed energy, wire consumed energy and switch consumed energy. To electric spark generation device, the efficiency of the device, $\varphi$, is the percentage of electric spark energy to released energy, that is $\varphi = E_{spark}/E_{release}$. The rate of electric spark energy release can be expressed by electric spark power, that is

$$P_{spark} = i^2 R_{spark} \tag{7}$$

## 2 Experimental results and analysis

### 2.1 Energy distribution of capacitor released energy

The electric discharge experiments were conducted by using the discharge system shown in Fig. 1. The electric spark with different energy level can be realized by selecting energy storage capacitor with different capacitance. For 10J, 100J and 1000 capacitor storage energy, the corresponding capacitance of energy storage capacitor are 400.5nF,

3.998μF, 31.34μF respectively.

To illustrate the details of energy calculation, a typical energy release process of 100J level storage energy was taken as an example. The charging voltage of the capacitor was about 11kV. And the voltage history of energy storage capacitor and the typical current history during capacitor discharge process were shown in Fig. 2.

By using the above mentioned method, the electric spark, wire and switch consumed energy can be calculated as follows:

$E_{spark} = 28.29J$, $E_{wire} = 50.64J$, $E_{switch} = 162.477J$

And the rate of energy conversion of the electric generation device is $\varphi = 11.7\%$.

**2.2 Results analysis**

2.2.1 The structure of energy distribution

The energy and the rate of energy release are the main parameters related to electric spark application in industry and scientific research. For an electric spark generated by capacitor discharge device, the characteristic parameters of electric spark are closely related to the circuit of device. During the energy release process of spark generation device, only a part of the capacitor released energy is transform into electric spark energy, and the others is consumed by wire and switch. In this section, the energy structure of capacitor released energy during capacitor discharge process was studied.

For the calculation of spark energy, three methods, that is *ui* integration method, Joule's law methods with spark resistance calculated by the parameters of current oscillation and over all energy conservation, were compared by Zhang Yunming et. al [7]. In their opinion, the Joule's law method with spark resistance calculated by the parameters of current oscillation had more advantages than the other two methods in calculating spark energy. In this paper, Joule's law had been used to evaluate electric spark energy, wire consumed energy and switch consumed energy with the resistance evaluated by energy conservation law and parameters of current oscillation respectively. The spark energy evaluated by Joule's law with resistance calculated by energy conservation law and parameters of current oscillation were denoted by $E_{spark1}$ and $E_{spark2}$ respectively. The capacitor released energy and the electric energy evaluated by two different methods described above with the capacitor stored energy level of 10J, 100J and 1000J were shown in Fig.3. Fig.3 showed that both of the released energy and the spark energy increased with the increase of initial charging voltage under different energy levels. The increasing rate of released energy with initial charging voltage is much greater than that of spark energy. The electric spark energy evaluated by Joul's law with resistance evaluated by energy conservation law follow conservation law strictly, that is, $E_{release} = E_{spark1} + E_{wire} + E_{switch}$, So, $E_{spark1}$ was used in analysis the energy structure of the released energy. And the higher the energy level is, the greater the proportion of spark energy to release energy goes, as shown in Fig. 3 and table 1.

The variations of electric resistance with charging voltage with different energy level were shown in Fig.4. From Fig. 4 we can see that with the increase of energy level, both $R_{total}$ and $R_{spark}$ decrease. The upper curve represents the total resistance, while the lower curve represents the spark resistance. Fig. 4 showed that the decreasing rate of $R_{total}$ is greater than that of $R_{spark}$. So, with the increasing of charging voltage at the same energy level and with the increasing of energy level, the ratio of spark resistance to total resistance increases, and the rate of energy conversion of the electric generation device increases. In addition, with the increasing of energy level, the capacitor released energy increased, the circuit is more prone to produce spark and the reliability of spark discharge has also been enhanced.

On the other hand, it can be found that $E_{spark2} > E_{spark1}$, and that can be explained by the different methods of spark resistance calculation. In the method of resistance calculation by the parameters of discharge current oscillation, the oscillation and attenuation of current is expressed approximately by exponentially damped sine function, and thus may result in some error. While in the method of resistance calculation by energy conservation law, current oscillation from experiment instead of approximation function is used. So the resistance calculated by energy conservation law is more accurate than that calculated by oscillation parameters of discharge current.

To study the structure of the released energy, the ratios of spark energy, wire consumption energy and switch consumption energy to total release energy have been calculated respectively, as shown in tab.1. It can be seen from tab. 1 that with the increasing of energy level, the percentage of spark energy to released energy and the percentage of wire consumed energy to released energy increased gradually while the percentage of switch consumed energy to released energy decreased gradually. With the increasing of energy level, the capacitor stored energy increase, the whole circuit is prone to discharge, the discharge current increased and the duration of discharging process becomes longer. However, the wire resistance can be seen as constant in a short time, so wire consumed energy becomes larger gradually. As shown in tab.1, switch resistance consumes most of the release energy.

The power of electric spark ($P_{spark}= i^2 R_{spark}$) is also an important parameter of electric spark discharge process. Fig.5 shows typical power history of electric spark discharge process under three different energy levels. It can be seen that the power of electric spark oscillated with amplitude exponentially damped. With the increase of energy level from 10J to 1000J, the period of power oscillation increased from 0.01ms to 0.05 ms, and the maximum power value is relatively increased from 700 kW to 1700kW. At lower energy level, the capacitor stored energy is small, and the duration of discharge process is short. While On the high energy level, the capacitor stored energy is large, the duration of discharge process is long. This phenomenon can be found both in histories of power, voltage and current.

## 3 Conclusions

1 The evaluation method for electric resistance during capacitor discharge process has been put forward based on energy conservation law. With the increasing of the capacitor stored energy from 10J to 1000J, the total resistance of discharge circuit decreased from 0.7 Ω to 0.08 Ω while the resistance of spark induced plasma decreased from 0.1 Ω to 0.012Ω.

2 The energy consumed by electric spark induced plasma, wire and switch have been calculated and the energy distribution structure of capacitor released energy has been analyzed. Both of the capacitor released energy and the spark energy increased with the increase of stored energy. In general, most of the capacitor released energy was consumed by circuit, including wire and switch. Just a little part of it (8.3%-13.5%) was transformed into electric spark energy. With the increasing of capacitor stored energy, the proportions of spark energy and wire consumed energy to capacitor release energy increased gradually. While the proportion of switch consumed energy to capacitor released energy decreased gradually.

3 The variation of power of electric spark induced plasma with time follows exponentially damped sinusoids function. With the capacitor stored energy increase from 10 J to 1000J, the maximum power of electric spark induced plasma increased from 700 kW to 1600kW and the duration of the discharge process increased from 0.06ms to 0.3 ms.

4 The histories of current and voltage of electric spark during the electric energy release process of capacitor have been measured and it was founded both voltage and current can be expressed as exponentially damped sinusoids function.

Figure and Table captions

Fig.1 Schematic graph of the spark discharge system

Fig.2 Typical histories of voltage and current during capacitor discharge process (100J—11KV)

Fig.3 Variations of energy with initial charging voltage

Fig.4 Variations of discharging resistance with initial charging voltage

Fig.5 Power history of electric spark

Table1 Energy distribution structure during capacitor discharge process

Figs and Tabs

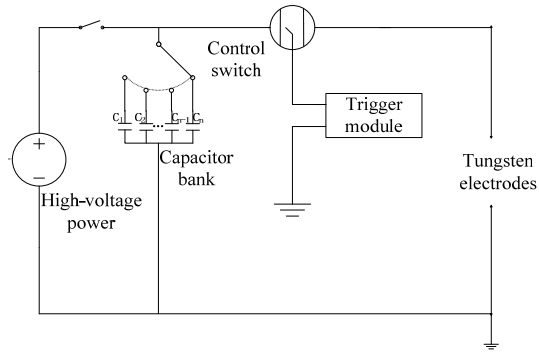

Fig.1

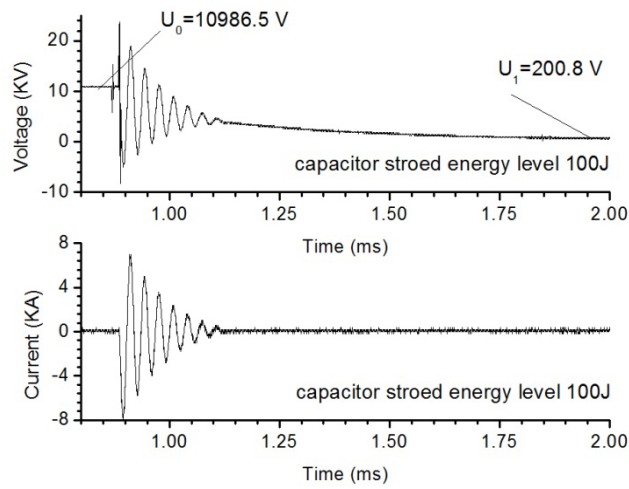

Fig.2

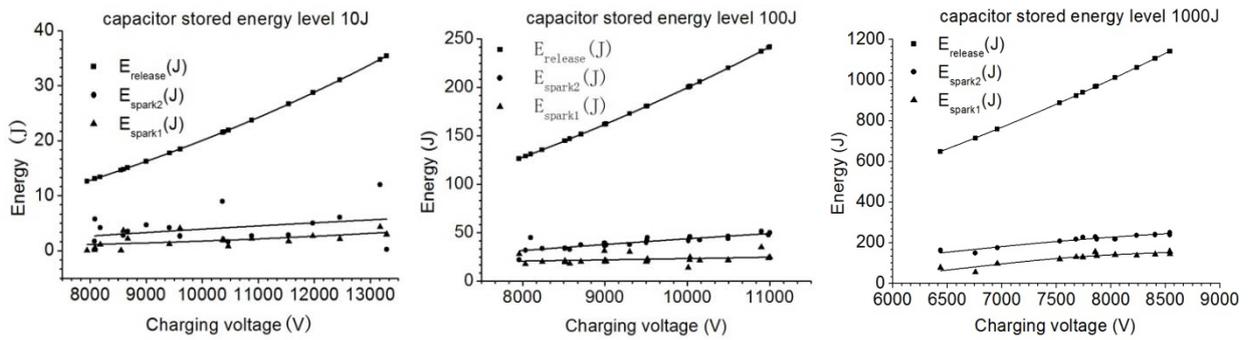

Fig.3

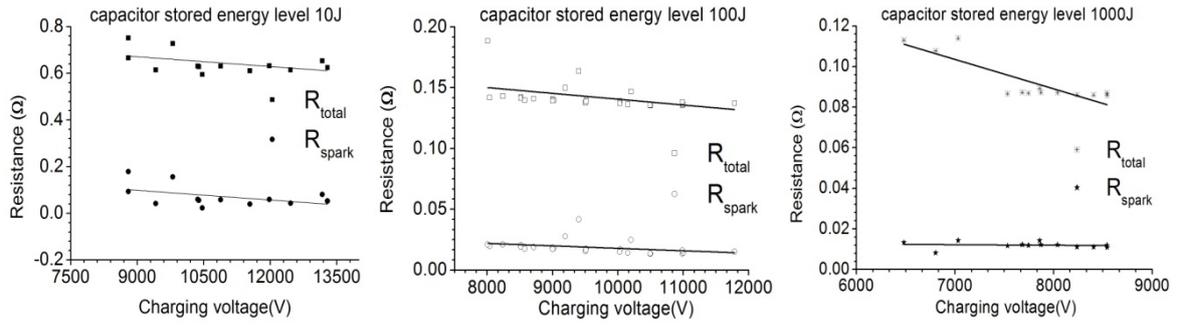

Fig.4

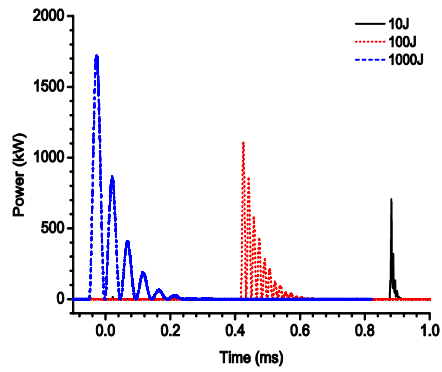

Fig.5

Table 1

| Energy level | $E_{release}$ | $E_{spark}$ | $E_{wire}$ | $E_{switch}$ |
| --- | --- | --- | --- | --- |
| 10J | 1 | 0.0830 | 0.0463 | 0.8672 |
| 100J | 1 | 0.1297 | 0.2056 | 0.6697 |
| 1000J | 1 | 0.1305 | 0.3207 | 0.5442 |